\documentclass[journal=jacsat,manuscript=article]{achemso}

\usepackage[version=3]{mhchem} 
\usepackage{lineno} 
\usepackage{xcolor}
\usepackage{amssymb} 
\usepackage{bm} 

\usepackage[shortcuts]{extdash} 

\author{Christoph Stockinger}
\affiliation[University of Graz]
{Institute of Physics, University of Graz, NAWI Graz, Universitätsplatz 5, 8010, Graz, Austria}
\alsoaffiliation[Christian Doppler Laboratory]
{Christian Doppler Laboratory for Structured Matter Based Sensing, Universitätsplatz 5, 8010, Graz, Austria}

\author{Natale G. Pruiti}
\affiliation[University of Glasgow]
{University of Glasgow, Rankine Building, Oakfield Avenue, Glasgow G12 8LT, UK}

\author{Isaac Tribaldo}
\affiliation[CSIC]
{Centro de Física de Materiales CSIC-UPV/EHU, Donostia-San Sebastián, Spain}

\author{Jörg S. Eismann}
\affiliation[University of Graz]
{Institute of Physics, University of Graz, NAWI Graz, Universitätsplatz 5, 8010, Graz, Austria}
\alsoaffiliation[Christian Doppler Laboratory]
{Christian Doppler Laboratory for Structured Matter Based Sensing, Universitätsplatz 5, 8010, Graz, Austria}

\author{Marc Sorel}
\affiliation[University of Glasgow]
{University of Glasgow, Rankine Building, Oakfield Avenue, Glasgow G12 8LT, UK}
\alsoaffiliation[Sant’Anna School of Advanced Studies, Pisa]
{Sant’Anna School of Advanced Studies, Via Moruzzi 1, 56124 Pisa, Italy}

\author{Peter Banzer}
\affiliation[University of Graz]
{Institute of Physics, University of Graz, NAWI Graz, Universitätsplatz 5, 8010, Graz, Austria}
\alsoaffiliation[Christian Doppler Laboratory]
{Christian Doppler Laboratory for Structured Matter Based Sensing, Universitätsplatz 5, 8010, Graz, Austria}
\email{peter.banzer@uni-graz.at}

\title[Passive Silicon Nitride On\-/Chip Polarimetry]
  {Passive Silicon Nitride On\-/Chip Polarimetry: Precise Polarization Detection with Imperfect Components}

\abbreviations{PIC,SEM,LCVR}

\begin{document}





\begin{abstract}
Polarization is a fundamental property of light that carries distinct and valuable information. Consequently, its precise measurement is crucial for numerous applications, including biomedical imaging, remote sensing, and optical communication. Since polarization cannot be measured directly, it is typically inferred by converting it into intensity signals using dedicated optical elements. Conventional approaches, however, predominantly rely on bulky optical components, leading to considerably high fabrication costs and limited integration density. Here, we introduce a passive photonic integrated circuit capable of precisely determining the polarization state of visible free\-/space light. An silicon nitride on\-/chip architecture employing a compact polarization\-/splitting grating coupler and a set of passive interferometers encodes the polarization information into intensity signals, allowing conventional detectors to accurately reconstruct the polarization state.  
With increasing compactness of photonic components, however, susceptibility to fabrication tolerances as well as intrinsic design constraints increases, potentially leading to non\-/ideal behaviour. To address this, we introduce a robust calibration procedure that enables precise measurements even in the presence of imperfections. The chip design, combined with the calibration procedure, offers a robust,
small\-/footprint, and high\-/speed approach to polarimetry, enabling a wide range of applications.
\end{abstract}

\section{Introduction}
The massive potential of photonic integrated circuits (PICs), offering unparalleled miniaturization, cost\-/efficient production, and chip\-/level integration, has positioned them at the forefront of photonics research.
With continuous advancements, PICs have found applications in various fields, including free-space light metrology \cite{Butow:22,Butow2023,Milanizadeh2022}.
Polarization, as one of the fundamental properties of light \cite{book_polarization}, is of particular interest in this context, and its detection is the main focus of this manuscript.
The precise measurement of polarization has proven great potential in various fields.
In biomedical imaging for example, it enables non\-/invasive diagnostics by detecting microscopic tissue changes, aiding in early disease detection \cite{vanLeeuwen2003,He2021,Pierangelo:11,https://doi.org/10.1002/jbio.201500006}.
Beyond this, polarization analysis is widely used in remote sensing \cite{DUBOVIK2019474}, optical communication \cite{Ip:08}, quantum communication \cite{RevModPhys.74.145} and other scientific and industrial sectors.

Polarization analysis has a long history, with traditional methods relying on bulky optical elements, such as in the rotating quarter-waveplate method \cite{10.1119/1.2386162} or liquid crystal-based techniques \cite{JuanMBueno_2000}.
Recent advances in meta\-/surfaces and plasmonics have enabled the development of compact polarimeters, eliminating the need for bulky optical components \cite{BalthasarMueller:16,https://doi.org/10.1002/lpor.201700297,Wei:17}.
Moreover, different waveguide-based on-chip polarimeter designs have been proposed, further enhancing integrability.
For instance, on\-/chip polarimetry can be performed using nanometer-scale scatterers \cite{doi:10.1021/acs.nanolett.7b00564, Espinosa-Soria2016} or inversely designed structures \cite{ZhouXieRenWeiDuZhangXieLiuLeiYuan+2022+813+819,WuYuLiuZhang+2019+467+474}, demonstrating promising performance while presenting challenges for large-scale fabrication.
Another approach involves directing two linearly polarized components of light under investigation into separate waveguides and analyzing them using either passive \cite{Lin:19,9021695} or reconfigurable \cite{ZhouZhaoWeiLiDongZhang+2019+2257+2267, 10.1063/1.5044379} on-chip interferometers.
Promising measurement results have also been reported for these methods. However, their integration is limited by conventional  polarization\-/splitting grating couplers, which act as free\-/space interfaces and polarization splitters but require long taper regions.
This limitation becomes particularly significant when considering two-dimensional detector arrays, for applications such as imaging polarimetry.
Notably, all waveguide-based polarimeters discussed are limited to the near-infrared spectral range.
However, recent advancements in reducing waveguide losses have made the development of complex, high-performance photonic integrated circuits for visible light applications possible \cite{Sacher:19,McKay_2023}.
Moreover, the compactness of photonic components, such as those used in the polarimeters mentioned above, not only offers advantages but also introduces susceptibility to manufacturing tolerances. Consequently, a calibration procedure following fabrication is often unavoidable \cite{Miller:15,Bandyopadhyay:21} and will be an important aspect of this work.

In this manuscript, we propose and experimentally demonstrate a passive silicon nitride photonic integrated circuit capable of detecting the polarization state of visible free-space light.
The design relies on a highly compact polarization-splitting grating coupler, engineered with focusing gratings.
The polarization state is analyzed on-chip using a fixed set of passive interferometers, which encode the complete polarization information into the output intensity.
The light is processed in a completely passive manner, enabling single\-/shot measurements.
As a result, detection time is constrained only by the response time of the intensity detectors.
Due to intrinsic design characteristics of the grating coupler and unavoidable fabrication tolerances, data evaluation requires a careful calibration process. 
To address this, we introduce a flexible calibration technique that enables highly accurate polarization measurements, even in the presence of imperfections in on-chip components.

\section{Results}

\subsection{Compact 2D Grating Coupler Design}

\begin{figure}[H]
\centering
\captionsetup{width=\linewidth}
\includegraphics[width = 0.5
\linewidth]{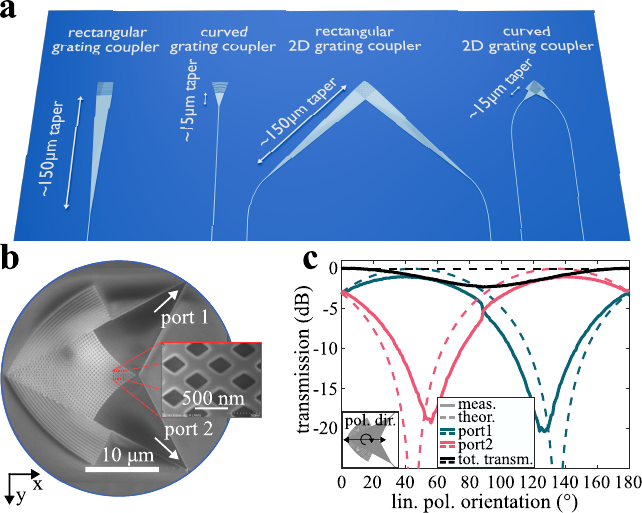}
\caption[2D Grating coupler]{Design of a compact two-dimensional grating coupler with a focusing grating design. (a) Schematic representation of different types of grating couplers. (b) SEM image of the curved two\-/dimensional grating coupler. (c) Measured (solid line) and ideal (dashed line) normalized transmission for the different ports of the grating coupler, with incident light linearly polarized at various orientations. The polarization direction was adjusted using the 2D polarization state generator shown in Fig. \ref{fig: setup}.}
\label{fig: 2d grating coupler}
\end{figure}

We start by introducing the free\-/space\-/to\-/chip interface used in our on\-/chip polarimeter. 1D grating couplers are commonly used in photonic integrated circuits to couple light from free\-/space into waveguides, with their grating patterns typically designed in either rectangular or curved geometries \cite{mi11070666}. Curved grating couplers offer focusing properties that reduce the need for extended taper regions, enabling a more compact component footprint 
\cite{VanLaere2007}. 
Although 1D grating couplers exhibit intrinsic polarization sensitivity due to waveguide birefringence, 2D grating couplers offer a more practical and versatile approach for applications resolving polarization \cite{8755389}. Ideally, these function as on-chip polarization beam splitters, positioned at the interface between free\-/space and on\-/chip circuitry, directing different polarization states into separate waveguides.
Furthermore, 2D grating couplers can also incorporate curved grating grooves to minimize taper regions, thus improving compactness \cite{4638145,8790808}. However, this design may compromise polarization splitting, which will be discussed in further detail below.
To demonstrate a size comparison of the different grating couplers, various grating coupler designs are depicted in a schematic illustration in Fig. \ref{fig: 2d grating coupler} (a).

A scanning electron microscope (SEM) image of the curved two-dimensional grating coupler used in this work is shown in Fig. \ref{fig: 2d grating coupler} (b).
The grating coupler is designed to couple the 45° linearly polarized component of the incident light into the fundamental TE mode of the waveguide at port 1, while the 135° polarized component is coupled into the fundamental TE mode of the waveguide at port 2.
Resulting from the compact design with minimized taper regions, the grating coupler was fabricated with a size of only 30 x 30 µm$^2$. 
While the curved grating provides significant benefits in terms of component size, it also introduces some technical challenges. 
Specifically, the overlap of curved grating grooves induces distortions in the grating voids, deviating from their usually rectangular shape, as shown in the inset of Fig. \ref{fig: 2d grating coupler} (b).
Hence, the chosen compactness of this key component of our on-chip polarimeter comes at the cost of imperfect polarization selectivity and extinction ratio, as discussed in more detail below.

Fig. \ref{fig: 2d grating coupler} (c) illustrates the normalized transmission for the different ports of the grating coupler, with incident light polarized at various orientations.
The dashed line represents the ideal behavior of a two-dimensional grating coupler, which perfectly splits the two orthogonal polarization components.
In contrast, the solid line shows the experimentally measured normalized transmission for the different ports of the compact two-dimensional grating coupler presented in this work.
The measurements reveal two significant deviations from ideal splitter of orthogonal linear polarization states.
First, the extinction ratio is substantially worse than in the ideal case. Hence, part of the incoming light is still coupled to port 1 or 2 even though the light is polarized orthogonally to the polarization direction that this port is supposed to be sensitive to.
Additionally, the measured transmission minima are shifted with respect to the ideal case. This indicates that the light coupled into the two ports is not purely 45° or 135° linearly polarized but instead consists of linearly polarized components with slightly different orientations that are not orthogonal to each other.
To describe the coupling characteristics of the curved 2D grating coupler, a simple theoretical model has been developed. This model will later enable us to circumvent the issues associated to the non\-/ideal polarization\-/splitting of curved 2D grating couplers in on\-/chip polarimeters through a calibration technique that we will discuss in detail below.

We can define a coupling matrix $\mathbf{\hat{C}}$ that relates the Jones vector of the free\-/space field to the waveguide modes:
\begin{equation}
    \begin{pmatrix}
      E^\mathrm{in}_1 \\
      E^\mathrm{in}_2
    \end{pmatrix} = \mathbf{\hat{C}}\begin{pmatrix}
      E_\mathrm{x} \\
      E_\mathrm{y}
    \end{pmatrix},
    \label{eq: transmission matrix - 1}
\end{equation}
where $E_\mathrm{x}$ and $E_\mathrm{y}$ represents the complex amplitudes of the x\-/ and y\-/polarized component of the free-space light, and $E^\mathrm{in}_1$ and $E^\mathrm{in}_2$ denote the complex electric field amplitudes of the waveguide modes at the two ports of the grating coupler.
The coupling matrix consists of two components; a projection matrix $\mathbf{\hat{P}}$, which accounts for the fact that the coupling does not occur strictly along the 45° or 135° polarization directions, and a cross\-/talk matrix $\mathbf{\hat{X}}$, which accounts for the cross\-/talk between the different polarization components:
\begin{equation}
    \mathbf{\hat{C}} = \mathbf{\hat{X}} \mathbf{\hat{P}}, \label{eq: transmission matrix - 2}
\end{equation}
with
\begin{align}
    \mathbf{\hat{P}} &= \begin{pmatrix}
            \cos(\alpha) & -\sin(\alpha)\\
            \cos(\alpha)& \sin(\alpha)
        \end{pmatrix}, \label{eq: transmission matrix - 3}\\
        \mathbf{\hat{X}} &= \begin{pmatrix}
        \sqrt{1-x_\mathrm{i}} & \sqrt{x_\mathrm{i}}\mathrm{e}^{\mathrm{i}\frac{\pi}{2}}\\
        \sqrt{x_\mathrm{i}}\mathrm{e}^{-\mathrm{i}\frac{\pi}{2}}& \sqrt{1-x_\mathrm{i}}
        \end{pmatrix}, \label{eq: transmission matrix - 4}
\end{align}
where $\alpha$ defines the effective polarization directions that the coupler channels into the two ports and $x_\mathrm{i}$ denotes the intensity cross-talk of the different polarization components. 

Note that coupling losses are not considered in this model. As a result, only the relative information about the coupled intensity is considered, which is sufficient for our intended application in an on-chip polarimeter.
%
%

\subsection{Polarimeter Design and Measurement Principle}
\begin{figure}[H]
    \centering
    \captionsetup{width=\linewidth}
    \includegraphics[width = \linewidth]{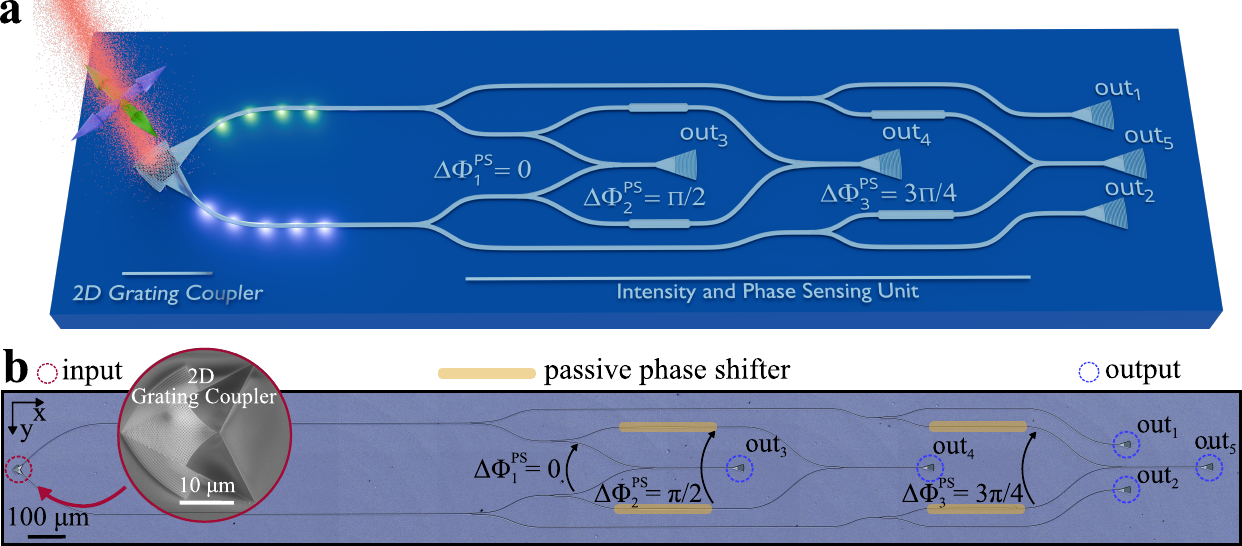}
    \caption[Optical microscopy image of the chip.]{
    Artistic illustration of the chip design (a) and optical microscopy image of the fabricated chip (b). A focusing 2D grating coupler serves as the interface that couples free-space light into the waveguides on the chip. This coupler splits the light into two linearly polarized components and directs each into separate waveguides. The two polarization components are then analyzed for their relative phase and amplitude using passive on-chip interferometers, providing complete information about the 2D polarization state of the incident light. Finally, the output signals of the on-chip architecture are coupled back into free space by means of standard grating couplers. To simplify experiments, the chip layout was intentionally designed with certain distances increased, resulting in a total footprint of 3295 × 325 µm.}
    \label{fig: microscopy image}
\end{figure}
Having introduced the two-dimensional grating coupler and its properties, we now integrate it into a photonic circuit to develop a fully passive on-chip polarimeter.
The two-dimensional grating coupler directs two linear polarization components of the incident light into separate waveguides, preserving both their relative amplitude and phase information.
As a result, the complete polarization information of the free-space light field is contained in the relative amplitude and phase of the waveguide modes. 
Recently, we introduced a method to retrieve the relative amplitude and phase between two waveguide modes via a single shot intensity measurement, using a set of all passive on-chip interferometers \cite{stockinger2024passivesiliconnitrideintegrated}.
We can thus integrate the two-dimensional grating coupler with the method described in Ref.~\citenum{stockinger2024passivesiliconnitrideintegrated} to develop a passive photonic chip capable of extracting polarization information from free-space light using only intensity measurements at the outputs.

A schematic of the photonic chip is shown in Fig. \ref{fig: microscopy image}(a).
The free\-/space\-/to\-/chip interface is realized using the 2D grating coupler introduced earlier (indicated in red in the microscope image of the chip in Fig. \ref{fig: microscopy image} (b)).  
A waveguide is connected to each port of the 2D grating coupler. These waveguides are then divided into several channels using Y-branch splitter \cite{Zhang:13}. 
Two of these channels are directed to output ports ($\mathrm{out_1}$ and $\mathrm{out_2}$, marked in blue in Fig. \ref{fig: microscopy image} (b)).
The outputs $\mathrm{out_1}$ and $\mathrm{out_2}$ provide direct information on the amplitude of the linearly polarized component being examined.
The remaining channels are paired using Y-branch combiner. 
In two of these pairs, the waveguides pass through a passive phase shifter before reaching the combiner. 
This phase shifter introduces a fixed relative phase shift between the waveguide modes, achieved simply by varying the waveguide width over a certain length \cite{Gonzalez-Andrade:20}.
Together with the passive phase shifters, the Y-branch combiners form passive on-chip interferometers that enable the effective retrieval of the relative phase in the input waveguides through intensity measurements at the chip's outputs.
At the outputs, the light is coupled back into free\-/space using standard 1D grating coupler, allowing the intensities to be measured via conventional detectors.

To gain a deeper understanding of the phase reconstruction, it is instructive to examine a theoretical model that connects the input fields on the chip (i.e., the fields in the waveguides after the 2D grating coupler)  with the output fields\cite{stockinger2024passivesiliconnitrideintegrated}:
\begin{align}
    E^\mathrm{out}_1 &= t_{11}E^\mathrm{in}_1, \label{eq: calib-1}\\ 
    E^\mathrm{out}_2 &= t_{22}E^\mathrm{in}_2, \label{eq: calib-2}\\
    E^\mathrm{out}_j &= t_{1j}E^\mathrm{in}_1 + t_{2j} E^\mathrm{in}_2~,~ \mathrm{for}~ j~\mathrm{ = ~3,~ 4,~ 5}.
    \nonumber\\
    &= A_{1j}\mathrm{e}^{\mathrm{i}\alpha_{1j}}+A_{2j}\mathrm{e}^{\mathrm{i}\alpha_{2j}},~ &\mathrm{for~} j \mathrm{~=~ 3,4,5}, 
    \label{eq: calib-3}
\end{align}
where $E$ represents the complex-valued electric field amplitude of the waveguide mode, while the complex proportionality coefficients $t_{ij}$ link the field at input i to the field at output j.
The amplitudes and relative phases of the coefficients $t_{ij}$ are determined through a calibration procedure, as outlined in a later section.
In \eqref{eq: calib-3}, we perform a substitution to separate complex\-/valued variables into real valued amplitude values $A_{ij} = |t_{ij}||E_\mathrm{in}^{i}|$ and their corresponding phase $\alpha_\mathrm{ij} = \tau_\mathrm{ij} + \phi_\mathrm{in}^{i}$, where $\tau_{ij} = \text{angle}(t_{ij})$ and $\phi_\mathrm{in}^{i} = \text{angle}(E_\mathrm{in}^{i})$.
Taking the modulus squared of \eqref{eq: calib-3}, using the relation $I \propto |E|^2$ and rearranging the equation leads to an expression for the relative phase:
\begin{equation}
    \alpha_{1j}-\alpha_{2j} = \pm \arccos\left(\frac{I^\mathrm{out}_j-(A_{1j}^2+A_{2j}^2)}{2A_{1j}A_{2j}}  \right) + 2\pi n, \label{eq: interferometer phase}
\end{equation}
with $\alpha_{1j}-\alpha_{2j} = \tau_{1j}-\tau_{2j}+\phi_1^\mathrm{in}-\phi_2^\mathrm{in}$, and $n$ an integer number.
We omit the term $2\pi n$ from our analysis, as it remains unresolved within our system and does not contribute relevant information to the polarization reconstruction.
Equation (\ref{eq: interferometer phase}) reveals that the relative phase $\alpha_{1j} - \alpha_{2j}$ can be determined, provided the output intensities of the interferometers and the amplitude factors $A_{1j}$ and $A_{2j}$ are known. Whereby the values of $A_{1j}$ and $A_{2j}$ can be derived using \eqref{eq: calib-1} and \eqref{eq: calib-2}, in conjunction with the intensities measured at the outputs $\text{out}_1$ and $\text{out}_2$, given the relationship  $I \propto |E|^2 $.
Finally, we need to address the challenge that \eqref{eq: interferometer phase} provides two possible signs in front of the inverse cosine function.
To determine the correct sign for a single interferometer, additional measurements are required that incorporate a known additional relative phase shift. In our case, this is accomplished using three passive interferometers equipped with distinct phase shifters.
The phase shifters are designed to introduce a fixed relative phase shift of $\Delta \Phi^\mathrm{PS}_1 = 0$, $\Delta \Phi^\mathrm{PS}_2 = \pi/2$ and $\Delta \Phi^\mathrm{PS}_3 = 3\pi/4$, providing balanced sensitivity of the interferometric phase reconstruction across all possible phase scenarios.
However, the phase shifts generated on the chip can deviate considerably from their design values due to minor manufacturing inaccuracies. Consequently, the actual phase shifts are determined through calibration and expressed in terms of the relative phases of the proportionality coefficients $t_{ij}$.

Once the relative amplitudes and phases of the waveguide modes are determined, the polarization state of the free\-/space light can be inferred, provided the behavior of the two-dimensional grating coupler is well understood.
Mathematically, this corresponds to rearranging equation \eqref{eq: transmission matrix - 1} by multiplying it from the left with the complex conjugate of the coupling matrix $\mathbf{\hat{C}}$.


\subsection{Experimental Setup}
\begin{figure}[H]
    \centering
    \captionsetup{width=\linewidth}
    \includegraphics[width = 0.5\linewidth]{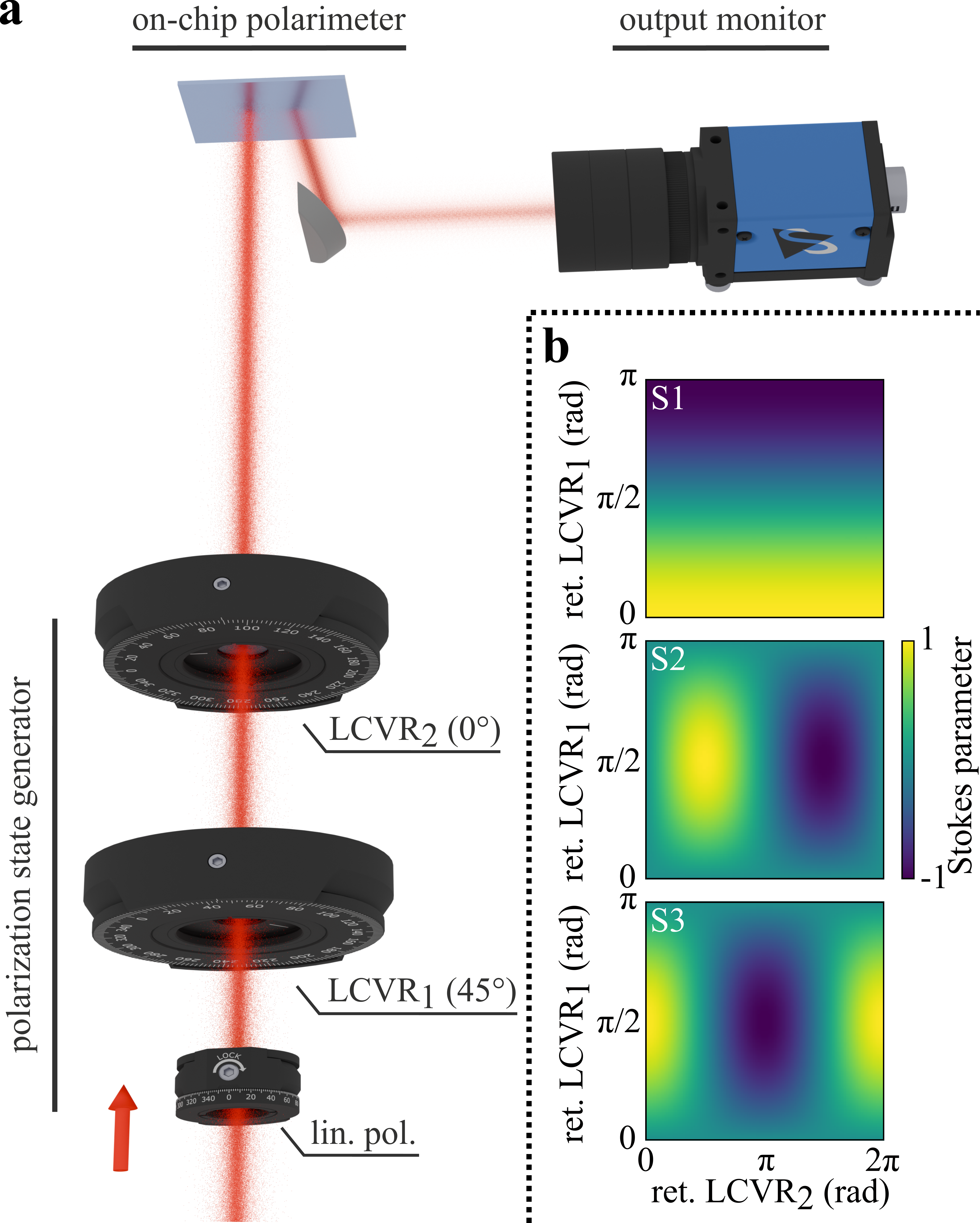}
    \caption[Illustration of the experimental setup.]{
    a) Illustration of the experimental setup: A collimated Gaussian beam first passes through a polarization state generator, which includes a horizontally aligned linear polarizer and two liquid crystal variable retarders (LCVRs) with their slow axes aligned at +45° and 0°, respectively. 
    The beam then enters the chip's input region. 
    The chip's output signals are monitored by means of an imaging system, consisting of a D\-/shaped mirror, an objective and a camera.
    b) Stokes parameters produced by the polarization state generator depending on the retardance set at the two liquid crystal variable retarders. The polarization state generator allows for the controlled creation of any possible polarization state by adjusting the retardance of LCVR$_1$ between 0 and $\pi$ and LCVR$_2$ between 0 and $2 \pi$.}
    \label{fig: setup}
\end{figure}

To enable precise calibration and facilitate a comprehensive investigation of the polarimeter's performance, it is crucial to design an experimental setup that offers full control over the polarization state of the light incident on the chip's input section, while simultaneously allowing accurate measurement of the chip's output signals.
A schematic of the key components of the setup used in this work is shown in Fig. \ref{fig: setup} (a).
The experimental setup employed for investigating the on\-/chip polarimeter begins with a fiber-coupled laser diode emitting light at a center wavelength of $\lambda = 658$ nm. 
We use a polarization state generator to set the polarization state of the light in a controlled manner. 
The polarization state generator consists of a horizontally aligned linear polarizer followed by two liquid crystal variable retarders (LCVRs), with their slow axes aligned at +45° and 0°, respectively. 
This configuration enables the generation of any desired polarization state by applying corresponding voltages to the liquid crystals.
The generated polarization states, based on the retardances set at the LCVRs, are shown in Fig. \ref{fig: setup} (b).
After the polarization state is set, the beam impinges onto the input region of the chip. 
The on\-/chip architecture encodes relative phase and intensity information from two linear polarization components of the beam into intensity signals. 
These output signals are monitored by an imaging system consisting of a mirror, an objective, and a camera.

\subsection{Calibration Technique}
Even minor manufacturing inaccuracies can alter the behavior of on-chip components. For instance, increased sidewall roughness of the waveguides can introduce losses in waveguide modes, while deviations in waveguide length or width can lead to unintended phase shifts \cite{641530}.
Furthermore, the design of specialized components can introduce unexpected and undesirable effects that impact their functionality, as seen in the case of the two-dimensional grating couplers with a focusing grating.
All these effects impact the chip's response to incident light with different polarizations and, if not accounted for, inevitably lead to increased measurement inaccuracies.
To assess the impact of hardware imperfections, a numerical error propagation simulation was conducted, the results of which are discussed in detail in the supplementary information.
Nonetheless, a robust calibration procedure can effectively compensate for these imperfections, allowing for highly accurate measurements even in the presence of significant imperfections in the on-chip components.
\begin{figure}[H]
    \centering
    \captionsetup{width=\linewidth}
    \includegraphics[width = \linewidth]{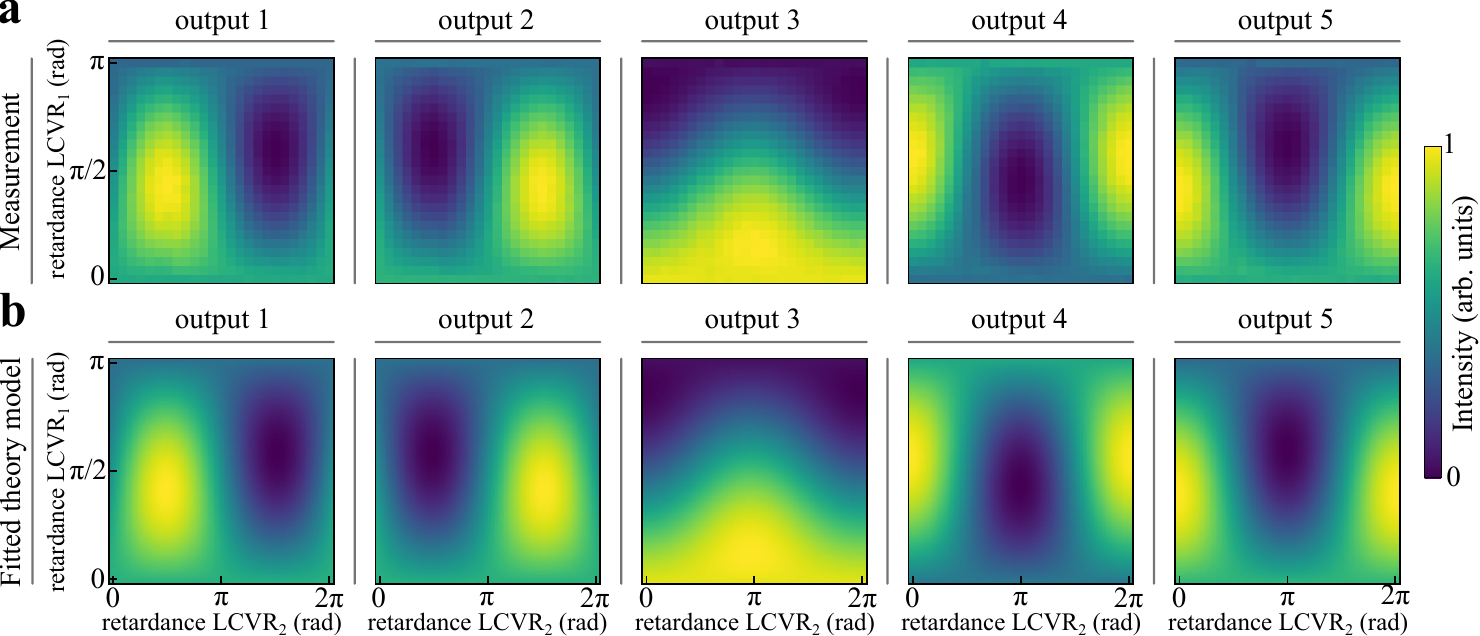}
    \caption[Results from the calibration procedure.]{Results from the calibration procedure, illustrating the output intensities of the chip for various input polarization states. 
    The input polarization states are produced by applying specific retardances to the liquid crystals of the polarization state generator, as depicted in Fig. \ref{fig: setup}. 
    Consequently, the intensities are presented as a function of the applied retardances. 
    (a) displays the experimentally measured output intensities, while (b) presents the theoretically calculated values obtained using the fitted theoretical model.
    }
    \label{fig: calibration}
\end{figure}
The calibration process involves illuminating the chip with light of precisely known parameters while simultaneously measuring the output signals. 
This allows us to determine the characteristics of the chip components.
To achieve this, we require a theoretical model of the chip that takes into account its imperfections.
We have already established this for the polarization-splitting grating coupler, which can be characterized by the coupling matrix $\mathbf{\hat{C}}$, as introduced in \eqref{eq: transmission matrix - 1}.
We have also developed a theoretical model for the remaining waveguide architecture, described by equations (\ref{eq: calib-1}) to (\ref{eq: calib-3}), which accounts for deviations arising from waveguide imperfections.
The model uses complex proportionality coefficients $t_{ij}$, whose amplitudes represent all potential losses experienced by the waveguide mode during propagation from input $i$ to output $j$.
The phases of the coefficients describe any relative phase shifts introduced to the waveguide mode, whether intentionally through passive phase shifters or unintentionally due to manufacturing inaccuracies.

Together with equations (\ref{eq: transmission matrix - 1}) to (\ref{eq: transmission matrix - 4}), which describe the coupling behavior of the polarization-splitting grating coupler, equations (\ref{eq: calib-1}) to (\ref{eq: calib-3}) form a comprehensive theoretical model that accounts for all relevant imperfections in the system, represented by the parameters $\alpha$, $x_\mathrm{i}$ and $t_{ij}$.

To determine the parameters, we use the setup shown in Fig. \ref{fig: setup} (a). 
The polarization state generator allows us to control the polarization of the reference beam, which we use for calibration.
To generate the broadest possible range of input polarization states, we vary the retardance of liquid crystal LCVR$_1$ from 0 to $\pi$ and the retardance of LCVR$_2$ from 0 to $2\pi$. This allows us to cover the entire surface of the Poincaré sphere, as shown in Fig. \ref{fig: setup} (b).
Simultaneously, we record the output intensities of the chip for each incident polarization state.
The experimentally obtained output intensities of the chip are presented in Fig. \ref{fig: calibration} (a).
A total of 625 distinct polarization states were generated by applying 25 x 25 different retardance values at the liquid crystals.
Since the generated input polarization states are precisely known, we can now extract the chip parameters from the measured output intensities.
This is achieved by fitting the previously discussed theoretical model of the chip to the experimental data.
The chip parameters $\alpha$, $x_\mathrm{i}$ and $t_{ij}$ are chosen as free variables in the fitting model, enabling them to be easily extracted once the algorithm converges.
While many fitting algorithms are suitable for this problem, we have chosen a standard least squares method.
The output intensities for the various input polarization states, calculated using the fitted theoretical model, are shown in Figure \ref{fig: calibration} (b).
The theoretical results closely match the experimental values.
Furthermore, the fitted parameters of the polarization-splitting grating coupler are consistent with the measurements presented in Fig. \ref{fig: 2d grating coupler} (b) and are further validated by FDTD simulations.
As previously observed in Ref. ~\citenum{stockinger2024passivesiliconnitrideintegrated}, owing to minor fabrication inaccuracies, the phase shifts introduced on the chip appear entirely random. However, since they can be accurately determined through the calibration procedure, they do not constitute a fundamental limitation to the proposed measurement method.

It is worth noting that the calibration process can be easily automated, as the reference beam is adjusted via the voltages applied to the liquid crystal retarders. 
This results in a rapid calibration procedure, enabled by the fast switching times of the liquid crystals, allowing the full calibration of a device to be completed in around two minutes.
\subsection{Polarization Measurements}
Once it is carefully calibrated, the photonic chip is capable of characterizing the polarization state of incident light with any arbitrary polarization.
A single shot intensity measurement of the chip's output signals is sufficient to deduce the polarization state of the light under investigation.
To extract the polarization information from the intensity data, the relative amplitudes and phases of the input waveguide modes are initially retrieved using the on-chip interferometric setup.
The obtained amplitude and phase information of the waveguide modes is subsequently translated into the Jones vector of the incident free-space light, using the coupling matrix $\mathbf{\Hat{C}}$ of the polarization-splitting grating coupler, which is determined during the calibration procedure.

\begin{figure}[H]
    \centering
    \captionsetup{width=\linewidth}
    \includegraphics[width = 0.5\linewidth]{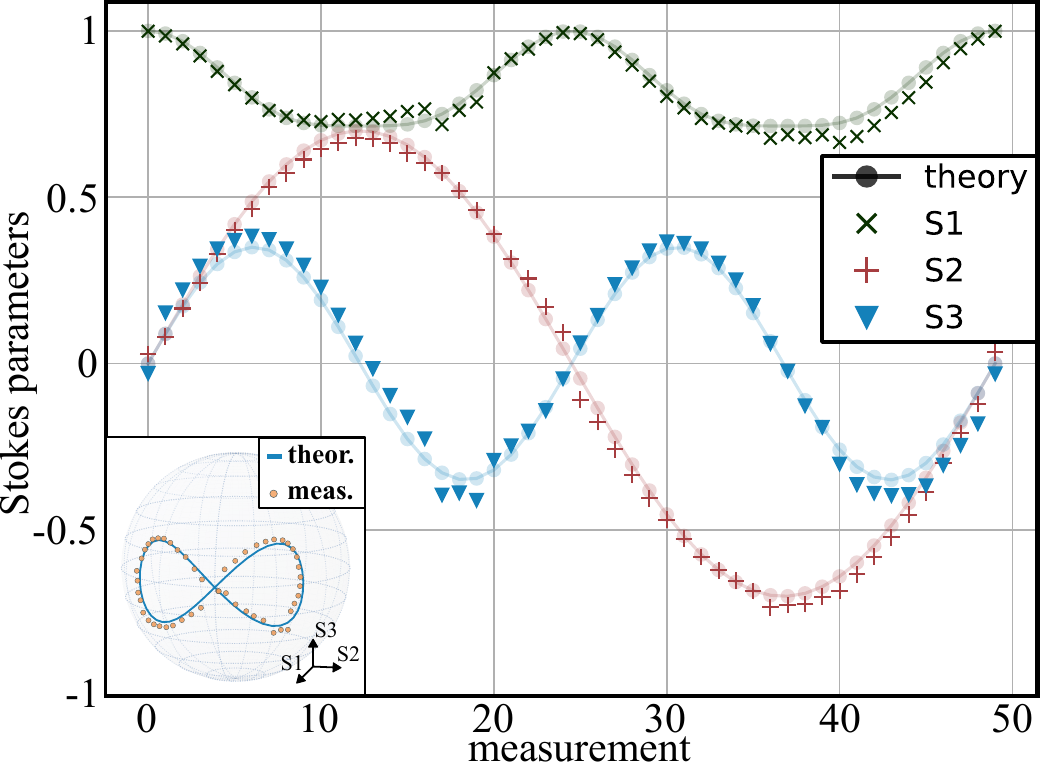}
    \caption[Full Stokes measurement]{
    Measured and theoretically expected Stokes parameters for 50 different polarization states, forming a closed loop on the surface of the Poincaré sphere. The distinct polarization states were probed by the input interface of the chip. The chip’s output intensities were used to reconstruct the polarization state of the input light. An alternative representation of the measurement results, with the measurement points depicted as dots on the surface of the Poincaré sphere, is displayed in the inset. The theoretical and experimental values show strong agreement, with a root mean square deviation of $\Delta S_{RMS} = 0.028$.}
    \label{fig: Stokes measurement}
    \end{figure}

To demonstrate polarization measurement using the photonic chip, we once again employ the experimental setup depicted in Fig. \ref{fig: setup} (a).
To evaluate the chip’s performance, we illuminate the input interface with light of 50 distinct polarization states, forming a closed loop on the surface of the Poincaré sphere.
The retrieved Stokes parameters for the different probed polarization states are shown in Fig. \ref{fig: Stokes measurement}.
The theoretical values correspond directly to the polarization states set at the polarization state generator.
There is strong agreement between the theoretical values and the experimentally determined values, with a root mean square deviation of $\Delta S_{RMS} = 0.028$ between the measured and theoretical Stokes parameters.

\section{Conclusion}
A photonic integrated circuit capable of detecting the polarization state of visible free-space light has been proposed and analyzed with respect to its performance.
A compact polarization-splitting grating coupler was introduced as the free-space-to-chip interface, employing a focusing grating design that results in a highly compact coupler size.
Subsequent to the grating coupler, the light processing on the chip is carried out through an architecture composed of a series of entirely passive interferometers.
This chip architecture enables the retrieval of the polarization state of incident free-space light through a single shot intensity measurement at the outputs.

It was found that the input grating coupler does not exhibit perfect polarization-splitting properties due to inherent characteristics of its design.
Additionally, the performance of other on-chip components may deviate from their ideal characteristics as a result of manufacturing inaccuracies.
To ensure accurate polarization measurements with the chip despite these imperfections, a straightforward, fast, and automatable calibration process was introduced.
Finally, we evaluated the calibrated chip's performance by probing 50 distinct polarization states.
Despite the presence of imperfect components, it was possible to achieve highly accurate polarization measurements.

It is important to note that advancements in integrated photonics such as on-chip detectors \cite{DeVita:22} or inverse\-/designed components \cite{Molesky2018} can be seamlessly integrated into our chip design, potentially leading to an even more compact and integrated device. 
Furthermore, possible extensions could include arranging multiple instances of the presented chip design in an array, enabling the effective measurement of the spatial distribution of the polarization of light fields.

The chip design, together with the methods presented, offers a powerful, cost-effective, and highly compact solution for on-chip polarimetry which is compatible with CMOS mass production.

\begin{acknowledgement}

The financial support by the Austrian Federal Ministry of Labour and Economy, the National Foundation for Research, Technology and Development and the Christian Doppler Research Association is gratefully acknowledged.
The financial support by EPSRC under project EP/X012689/1 is gratefully acknowledged. The authors acknowledge the technical staff at the James Watt Nanofabrication Centre (JWNC) at the University of Glasgow.
The authors acknowledge Gabriel Molina-Terriza for valuable comments on the manuscript and for proofreading.

\end{acknowledgement}


\bibliography{literature}
\appendix

\section{Supplementary Information}

\setcounter{equation}{0}
\renewcommand{\theequation}{S\arabic{equation}}

\setcounter{section}{0}
\renewcommand{\thesection}{S\arabic{section}}

\setcounter{figure}{0}
\renewcommand{\thefigure}{S\arabic{figure}}

\setcounter{table}{0}
\renewcommand{\thetable}{S\arabic{table}}
\subsection{Numerical polarization measurement error analysis}

There are several potential sources for measurement errors when determining the polarization state of free-space light using the polarimeter proposed in the main text. These range from inherent photodetection noise, which is unavoidable in practice, to imperfections in the hardware of on-chip components. In this section, we present a numerical analysis of the various error mechanisms affecting our system.

One effective way to evaluate the impact of various error sources on measurement accuracy is to perform statistical simulations with random sampling, as demonstrated in Ref. \cite{Bandyopadhyay:21}. In this approach, the input parameters of the chip’s theoretical model are randomly varied while applying random or systematic perturbations to the system. The resulting outputs are subsequently analyzed using statistical methods to evaluate the impact of these perturbations on the overall measurement accuracy.

We use Equations (1) to (7) from the main text to simulate the polarimeter. This model is fed with input fields (representing input polarizations) of the form:
\begin{align}
      E_\mathrm{x,y} \propto \frac{1}{\sqrt{2}}\left(\mathcal{N}(0,1)+\mathrm{i}\mathcal{N}(0,1)\right), 
    \label{eq: input fields}
\end{align}
where $\mathcal{N}(0,1)$ is the normal distribution with mean zero and standard deviation one.
For specific error scenarios, we calculate the expectation of the measurement error for each Stokes parameter S1, S2, and S3 to characterize the average error in polarization reconstruction:

\begin{align}
\mathbb{E}[|\Delta \mathrm{S}|] \approx \frac{1}{N} \sum_{i=1}^{N} |\Delta \mathrm{S}_i|
\end{align}

where $N$ is the total number of simulations and $\Delta \mathrm{S}_i = \mathrm{S}_{i,\text{sim}} - \mathrm{S}_{i,\text{true}}$.
To ensure reliable statistics, we perform $N=10^5$ simulations for each scenario.

In the following, we will examine and discuss the simulations described above, considering the effects of photodetection noise, polarization projection deviation of the 2D grating coupler, polarization crosstalk in the 2D grating coupler, phase errors in the on-chip phase shifters, and random unbalanced propagation losses in waveguides.

\subsubsection{Photodetection noise}
We start with an analysis of the measurement error introduced by photodetection noise. To this end, a simplified noise model is employed, where Gaussian noise is added to the simulated output intensities of the polarimeter:
\begin{align}  
I^{\text{out}\bm{'}}_i = I^{\text{out}}_i + \sigma_n \mathcal{N}(0,1),
\end{align}  
where $\sigma_n=\frac{1}{SNR}$.

\begin{figure}[H]
    \centering
    \captionsetup{width= \linewidth}
    \includegraphics[width = \linewidth]{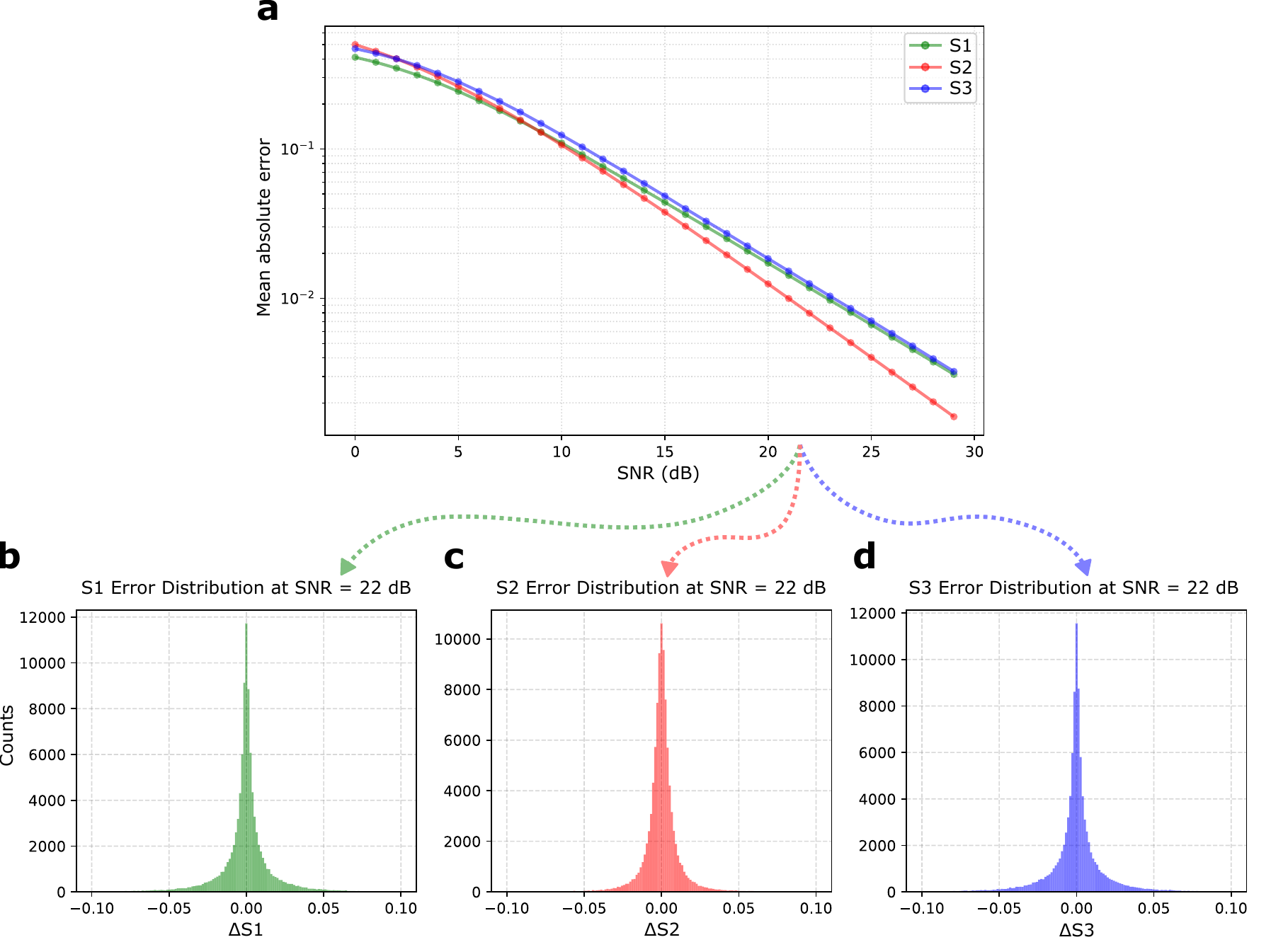}
    \caption[Error estimation for photo detection noise.]{Simulated polarization sensing errors analyzed under varying photodetection noise conditions. (a) shows the mean absolute error (averaged over $10^5$ simulations) for the Stokes parameters as a function of the simulated signal-to-noise ratio (SNR). (b) to (d) present histograms of the Stokes parameters at a signal-to-noise ratio (SNR) of 22 dB, generated from $10^5$ randomly sampled input polarizations.
    }
    \label{fig: error_estimate_pdnoise}
\end{figure}
The simulated polarization measurement error under varying photodetection noise conditions is presented in Figure \ref{fig: error_estimate_pdnoise}. 
Figure \ref{fig: error_estimate_pdnoise} (a) illustrates the mean absolute measurement error of the Stokes parameters as a function of the signal-to-noise ratio (SNR). 
Figure \ref{fig: error_estimate_pdnoise} (b) to (d) display histograms of simulated measurement errors generated with random photodetection noise at an SNR of 22 dB, which corresponds to the estimated photodetection noise for the measurements discussed in the main text.
The results indicate that the mean absolute measurement error is inversely proportional to the SNR, except at low SNR values. Additionally, for gaussian photodetection noise, the measurement errors exhibit a symmetrical distribution.

Given that photodetection noise is inherent in experiments, a Gaussian photodetection noise with an SNR of $ \sim$ 22 dB is consistently incorporated into the error model when analyzing other sources of error.
\subsubsection{Imperfect polarization projection}

\begin{figure}[H]
    \centering
    \captionsetup{width=\linewidth}
    \includegraphics[width = \linewidth]{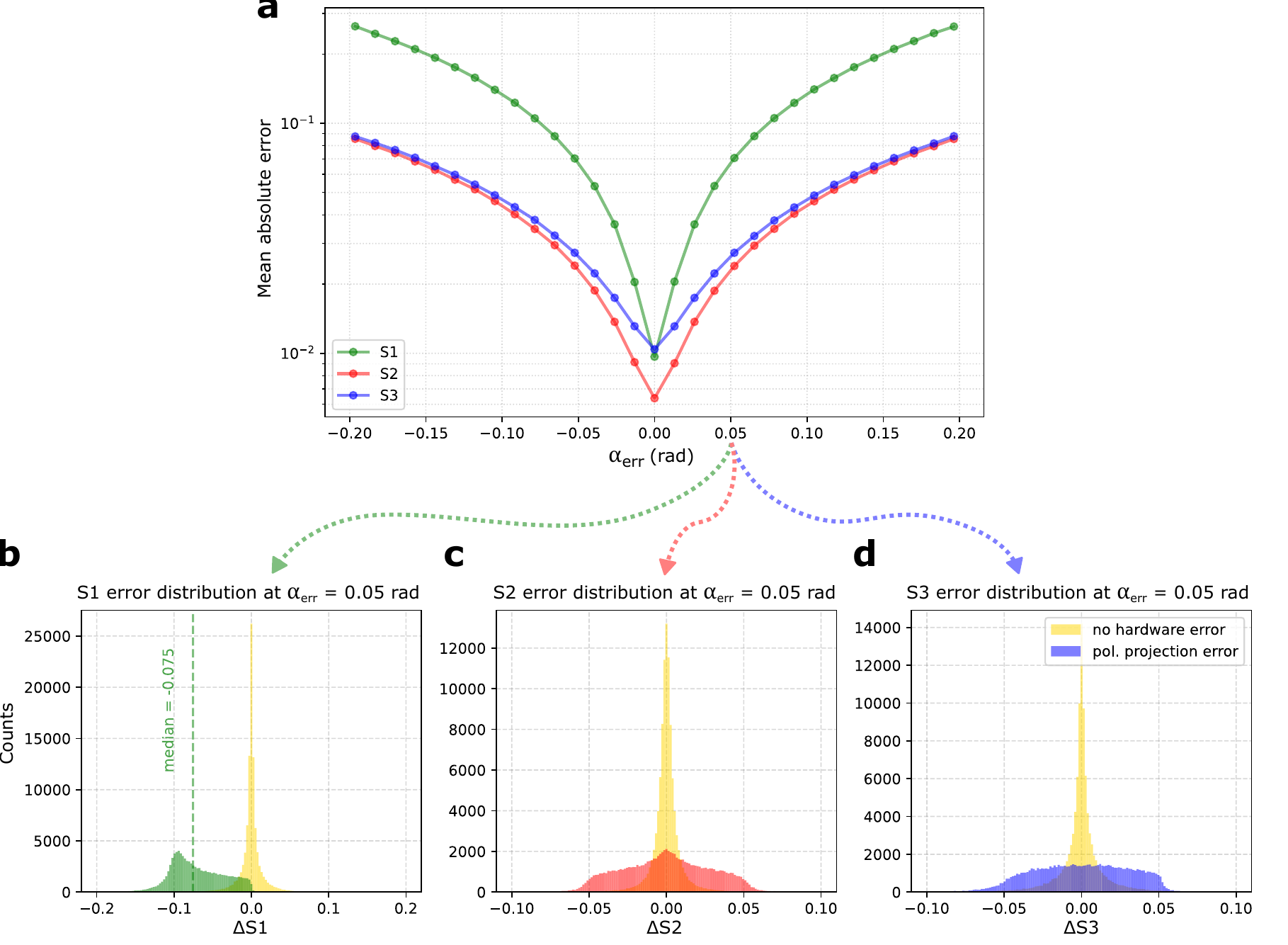}
    \caption[Error estimation for polarization projection error.]{Simulated polarization sensing errors analyzed under varying polarization projection hardware error. (a) shows the mean absolute error (averaged over $10^5$ simulations) for the Stokes parameters as a function of varying polarization projection deviation $\alpha_\mathrm{err}$. (b) to (d) show histograms of the $10^5$ simulated measurement errors for the Stokes parameters at $\alpha_\mathrm{err} = 0.05$ rad (green, red and blue)  and in the absence of hardware error (yellow).
    }
    \label{fig: error_estimate_pol_projection}
\end{figure}

Next, we examine deviations in the polarization projection during the coupling process using 2D grating couplers. The polarization projection performed by the coupler is described by the projection matrix $\mathbf{\hat{P}}$, as presented in equation (3) of the main text. In the following, we introduce a biased deviation from the ideal projection, defined as follows:
\begin{align}
    \alpha = \frac{\pi}{4}+\alpha_\mathrm{err}.
\end{align}
The simulated polarization measurement error under varying polarization projection deviation is presented in Figure \ref{fig: error_estimate_pol_projection}. 
Figure \ref{fig: error_estimate_pol_projection} (a) illustrates the mean absolute measurement error of the Stokes parameters as a function of the polarization projection error. 
Figure \ref{fig: error_estimate_pol_projection} (b) to (d) display histograms of the simulated measurement errors generated with a biased polarization projection error of $\alpha_\mathrm{err} = 0.05$ rad.
The results reveal several notable effects. Firstly, the measurement error increases significantly with larger polarization projection errors. Additionally, a biased measurement error is observed, particularly for the Stokes parameter S1, which exhibits a median value of $ \sim-0.075$ at a projection error of $\alpha_{err} = 0.05$ rad.

\subsubsection{Polarization cross-talk}
The next source of error we examine is polarization cross-talk, which can occur during the polarization splitting process in 2D grating couplers. The polarization cross-talk is mathematically represented by the cross-talk matrix $\mathbf{\hat{X}}$ in equation (4) of the main text, where $x_\mathrm{i}$ quantifies the cross-talk. Ideally, $x_\mathrm{i} = 0$, indicating the absence of polarization cross-talk. In our error model, we systematically vary $x_\mathrm{i}$ to analyze its impact on measurement errors.

\begin{figure}[H]
    \centering
    \captionsetup{width=\linewidth}
    \includegraphics[width = \linewidth]{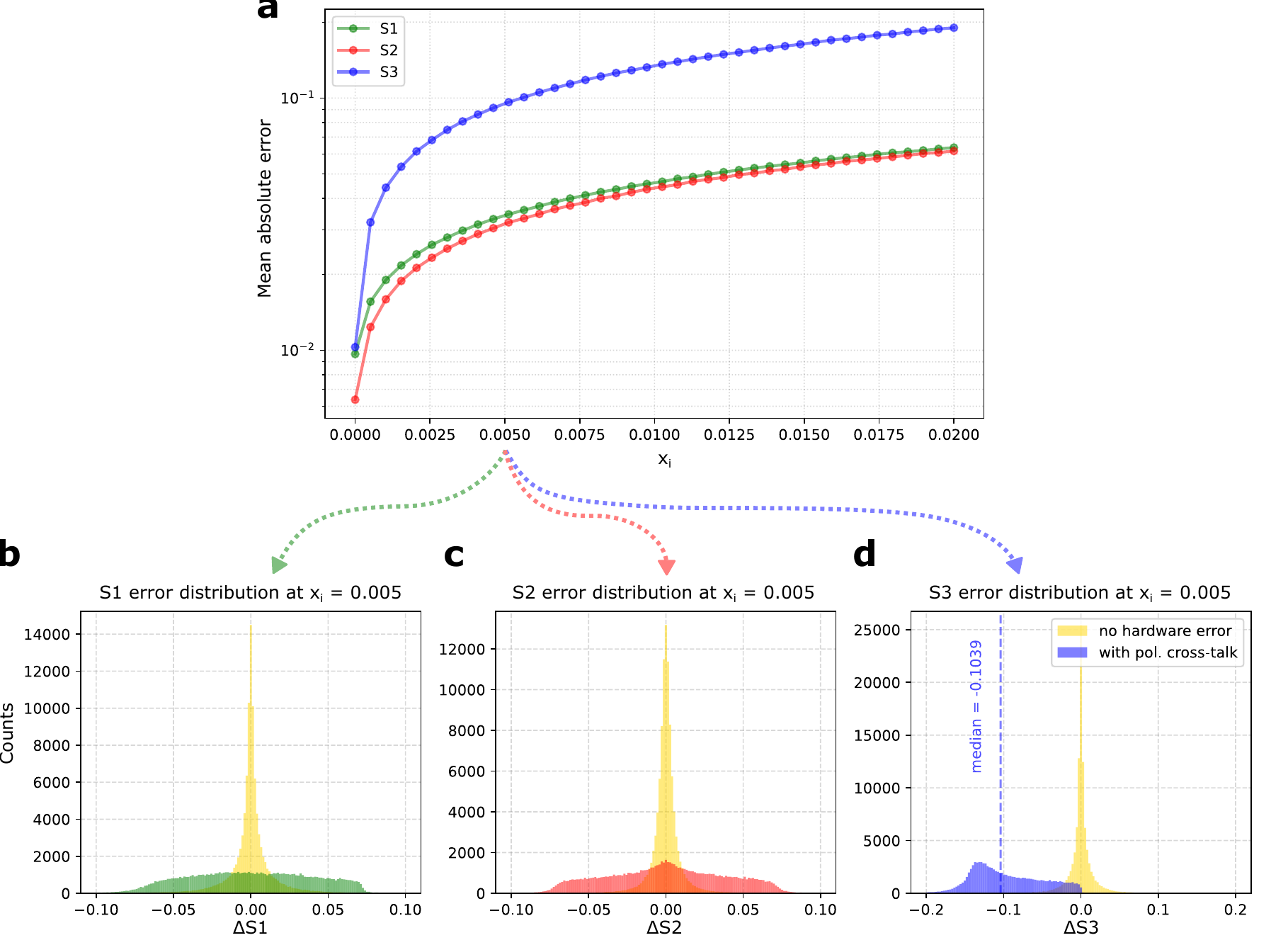}
    \caption[Error estimation for polarization cross-talk.]{Simulated polarization sensing errors analyzed under varying polarization cross-talk $x_\mathrm{i}$. (a) shows the mean absolute error (averaged over $10^5$ simulations) for the Stokes parameters as a function of the polarization cross\-/talk. (b) to (d) show histograms of the $10^5$ simulated measurement errors for the Stokes parameters at $x_\mathrm{i} = 0.005$ (green, red and blue)  and in the absence of hardware error (yellow).
    }
    \label{fig: error_estimate_pol_cross_talk}
\end{figure}
The simulated polarization measurement error under varying polarization cross\-/talk is presented in Figure \ref{fig: error_estimate_pol_projection}. 
Figure \ref{fig: error_estimate_pol_projection} (a) illustrates the mean absolute measurement error of the Stokes parameters as a function of the polarization cross\-/talk. 
Figure \ref{fig: error_estimate_pol_projection} (b) to (d) show histograms of the simulated measurement errors obtained for a polarization cross\-/talk of $x_\mathrm{i} = 0.005$.
The analysis shows that the measurement error increases significantly with higher levels of polarization cross\-/talk. Additionally, a biased measurement error is observed for the Stokes parameter S3, which exhibits a median value of approximately -0.104 for a polarization cross\-/talk of $x_\mathrm{i} = 0.005$.

\subsubsection{Imperfect phase shifters}

The passive phase shifters implemented in the on\-/chip interferometers represent additional elements that can deviate from their nominal design specifications due to manufacturing imperfections.
To investigate the impact of phase shifter variations on the measurement accuracy of the polarimeter, random deviations are introduced to the relative phase shifts imposed by the phase shifters:
\begin{align}
    \Delta\Phi_i^\mathrm{PS\bm{'}} =    \Delta\Phi_i^\mathrm{PS}+\sigma_\mathrm{PS} \mathcal{N}(0,1).
\end{align}

\begin{figure}[H]
    \centering
    \captionsetup{width=\linewidth}
    \includegraphics[width = \linewidth]{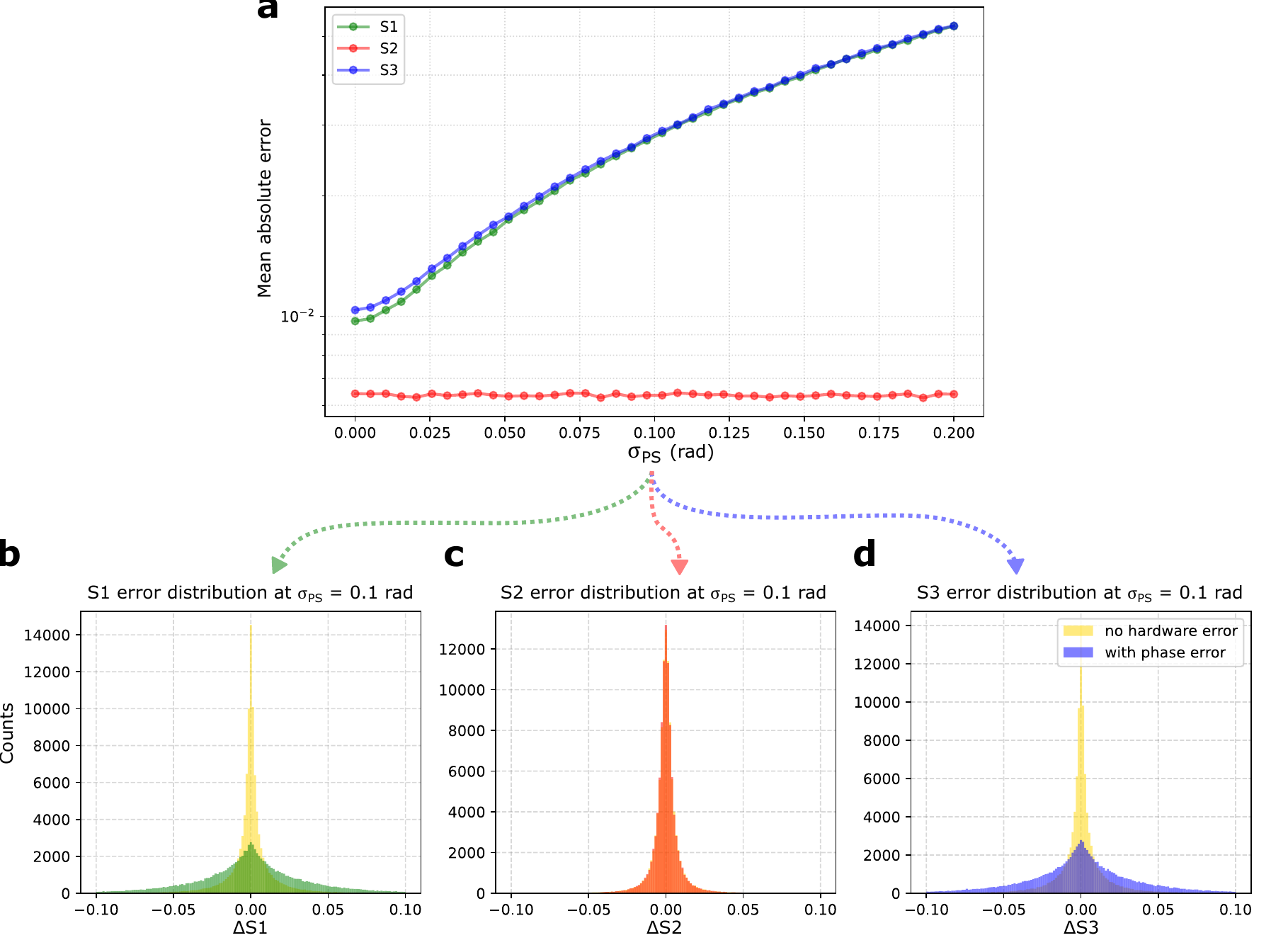}
    \caption[Error estimation for phase shifter deviation.]{Simulated polarization sensing errors analyzed under varying phase shifter errors $\sigma_\mathrm{PS}$. (a) shows the mean absolute error (averaged over $10^5$ simulations) for the Stokes parameters as a function of the phase error. (b) to (d) show the histograms of the simulated measurement errors for the Stokes parameters at $\sigma_\mathrm{PS} = 0.1$ rad (green, red and blue)  and in the absence of hardware error (yellow), based on $10^5$ randomly generated input polarizations.
    }
    \label{fig: error_estimate_PS}
\end{figure}
Figure \ref{fig: error_estimate_PS} (a) illustrates the mean absolute measurement error of the Stokes parameters as a function of the phase shifter error. 
Figure \ref{fig: error_estimate_PS} (b) to (d) depict histograms of the simulated measurement errors obtained for a phase error of $\sigma_\mathrm{PS} = 0.1$ rad.
The error analysis indicates that the measurement errors of S1 and S3 increase with the magnitude of the phase error, whereas S2 remains unaffected.

In addition to random phase shift errors caused by manufacturing tolerances, systematic errors can also arise, such as those stemming from consistent over- or under-etching during the fabrication process. Our error analysis reveals that the overall behavior of biased phase errors is comparable to that of random phase errors. Therefore, this source of error will not be further elaborated on.
\subsubsection{Random propagation loss}
Finally, we investigate the effects of random propagation losses in waveguides, characterized by the amplitudes of the proportionality coefficients $t_{ij}$. To evaluate waveguide losses, the amplitudes of the coefficients are randomly perturbed as follows:

\begin{align}
    |t_{ij}^{\bm{'}}| = |t_{ij}|+\sigma_\mathrm{t}\mathcal{N}(0,1).
\end{align}

\begin{figure}[H]
    \centering
    \captionsetup{width=\linewidth}
    \includegraphics[width = \linewidth]{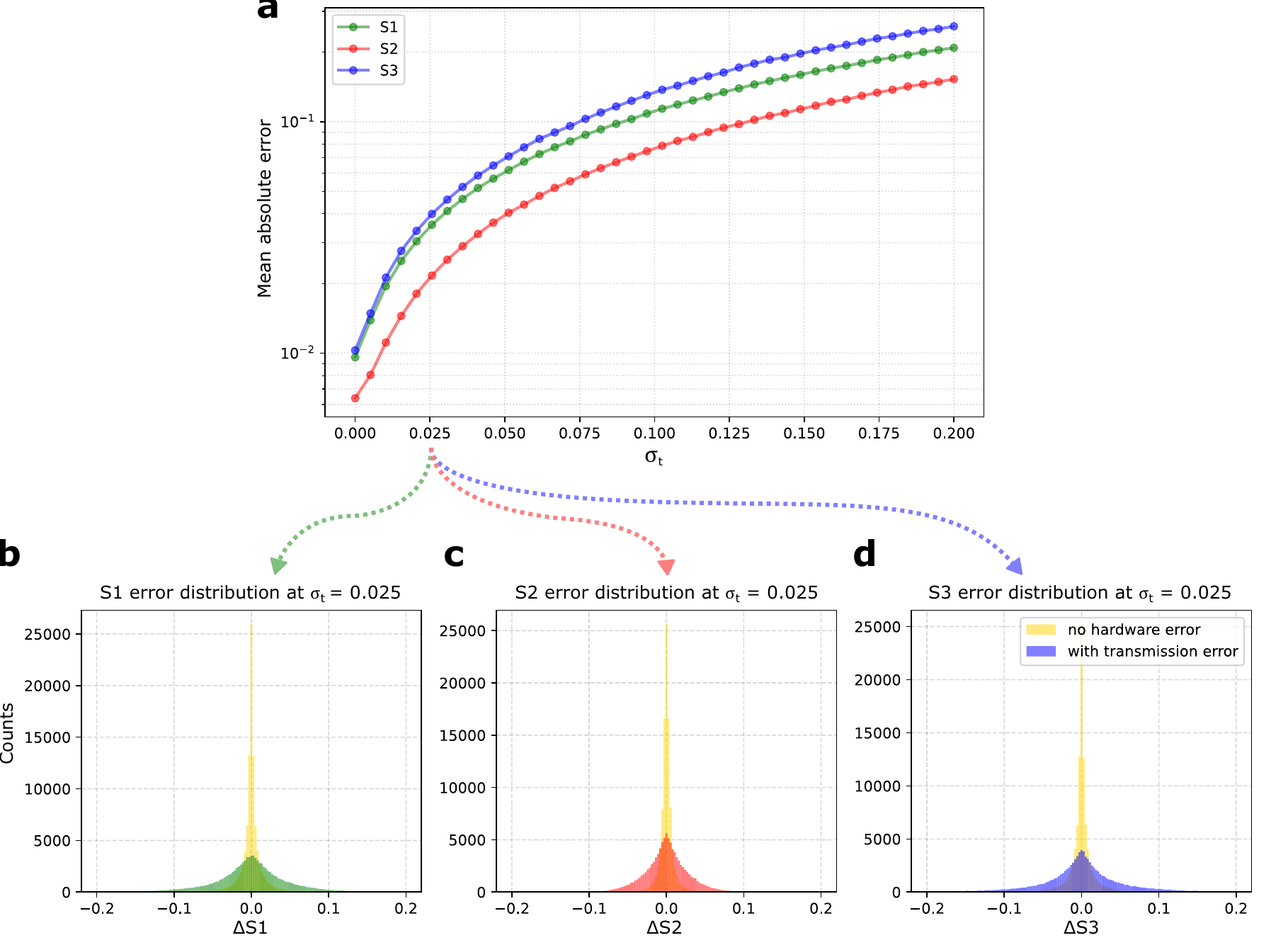}
    \caption[Error estimation for random propagation loss.]{Simulated polarization sensing errors analyzed under varying propagation loss. (a) shows the mean absolute error (averaged over $10^5$ simulations) for the Stokes parameters as a function of $\sigma_\mathrm{t}$. (b) to (d) show histograms of the $10^5$ simulated measurement errors for the Stokes parameters at $\sigma_\mathrm{t} = 0.025$ rad (green, red and blue)  and in the absence of hardware error (yellow).
    }
    \label{fig: error_estimate_prop_noise}
\end{figure}

Figure \ref{fig: error_estimate_prop_noise} (a) illustrates the mean absolute measurement error of the Stokes parameters as a function of $\sigma_\mathrm{t}$. 
Figure \ref{fig: error_estimate_PS} (b) to (d) depict histograms of the simulated measurement errors obtained for a phase error of $\sigma_\mathrm{t} = 0.025$.
The error analysis reveals that the measurement error increases for all three Stokes parameters as random propagation losses increase.

In this error scenario, it is important to note that biased propagation losses may occur as a result of the chip architecture, wherein light propagates through waveguides of different lengths before reaching the outputs. However, to maintain simplicity, this error is excluded from the analysis.

\end{document}